\documentclass[12pt,aps,floats,showpacs,amssymb,tightenlines]{revtex4}
\usepackage{color}
\usepackage{graphicx}
\usepackage{amsmath}
\usepackage{amsfonts}
\usepackage{amscd}
\usepackage{epsfig}
\usepackage{amssymb}
\usepackage{tabularx}

\begin{document}

\title{Static spherically
symmetric Einstein-Vlasov shells made up of particles with a discrete set of values of their angular momentum}
\author{Reinaldo J. Gleiser} \email{gleiser@fis.uncor.edu} \author{Marcos A. Ramirez}
\affiliation{Facultad de Matem\'atica,
Astronom\'{\i}a y F\'{\i}sica, Universidad Nacional de C\'ordoba,
Ciudad Universitaria, (5000) C\'ordoba, Argentina}

\begin{abstract}
In this paper we study static spherically
symmetric Einstein-Vlasov shells, made up of equal mass particles, where the angular momentum $L$ of particles takes values only on a discrete finite set. We consider first the case where there is only one value of $L$, and prove their existence by constructing explicit examples. Shells with either hollow or black hole interiors have finite thickness. Of particular interest is the thin shell limit of these systems and we study its properties using both numerical and analytic arguments to compare with known results. The general case of a set of values of $L$ is also considered and the particular case where $L$ takes only two values is analyzed, and compared with the corresponding thin shell limit already given in the literature, finding good agreement in all cases.
\end{abstract}
\pacs{04.50.+h,04.20.-q,04.70.-s, 04.30.-w}

\maketitle

\section{Introduction}

Although sometimes sidestepped, it is a general requirement in studying non vacuum spacetimes in general relativity that the energy momentum tensor, that is the matter (field) contents, should have a clear, although possibly highly idealized, physical interpretation. Among these choices the case where matter is described as a large ensemble of particles that interact only through the gravitational field that they themselves, at least partially, create, is of particular interest, both because of their usefulness in modeling physical systems such as star o galaxy clusters, and of the possibility of a relatively detailed analysis, at least in some restricted cases. As usual in theoretical treatments, one starts imposing as many restrictions as compatible with the central idea, and then tries to generalize from these cases. In this respect, the restriction to static spherically symmetric systems provides an important  simplification, although even with this restriction the problem is far from trivial, and further restrictions have been imposed in order to make significant advances. One of the first concrete examples is that provided by the Einstein model \cite{einstein}, where the particles are restricted to move on circular orbits. This model is static, and it is not easy to generalize as such to include a dynamical evolution of the system. This generalization can, however, be achieved if the particle world lines are restricted to a shell of vanishing thickness( ``thin shell''), as considered by Evans in \cite{Evans}. The analysis in \cite{Evans}, although, motivated by the Einstein model, considers only shells where all the component particles have the same value of their (conserved) angular momentum. In a recent study \cite{gleram} of the dynamics of spherically symmetric thin shells of counter rotating particles, of which \cite{Evans} is an example, it was found that the analysis can be extended to  shells where the particles have angular momenta that take values on a discrete (but possibly also continuous) set, and is not restricted to a single value. It was also found that in the non trivial thin shell limit of a thick Einstein shell the angular momentum of the particles acquires a unique continuous distribution, and, therefore, the models in \cite{Evans} and \cite{gleram} are not approximations to an Einstein model. A relevant question then is what, if any, are the (thick) shells that are approximated by those in \cite{Evans} and \cite{gleram}. In this paper we look for an answer to this question by considering a generalization of the Einstein model where instead of circular orbits we impose, at first, the restriction to a single value of the angular momentum. The particle contents is described by a distribution function $f$ in phase space, and, because of the assumption of interaction only through the mean gravitational field, $f$ must satisfy the Einstein-Vlasov equations \cite{refEV}. In the next Section we set up the problem and show that it leads to a well defined set of equations. In Section III we set up and analyze a particular model, obtaining expansions for the metric functions at the boundary of the support of $f$, appropriate for numerical analysis. Further properties are analyzed in Section IV, where we show that all these shells have finite thickness. Section V contains numerical results for a generic example. The   ``thin shell'' limit is considered in Section VI, both through analytic arguments and a concrete numerical example, with the results showing total agreement with the thin shell results of \cite{gleram}. A further comparison with \cite{gleram} is carried out in Section VII, where the stability of a shell approaching the thin shell limit is considered. The generalization to more than one value of $L$ is given in Section VIII, where we find that particles with different values of $L$ may be distributed on shells that overlap completely, or do so only partially or not at all. Numerical examples and a comparisons with \cite{gleram} are finally developed in Section IX. Some comment and conclusions are given in Section X.

\section{ The static spherically symmetric Einstein-Vlasov system}

The metric for a static spherically symmetric spacetime may be written in the form,
\begin{equation}\label{one1}
ds^2= -B(r) dt^2 + A(r) dr^2 +h(r)^2 d \Omega^2
\end{equation}
where $d \Omega^2 =d\theta^2 +\sin^2 \theta d\phi^2$ is the line element on the unit sphere and $r \geq 0$.

For a static, spherically symmetric system, the matter contents, in this case equal mass collisionless particles, is microscopically described by a distribution function $F(r,p^j)$, where $p^j=(p^r,p^{\theta},p^{\phi})$ are the components of the particle momentum, taken per unit mass. Then, as a consequence of the assumption that the particles move along geodesics of the space time metric, the distribution function $F$ satisfies the Vlasov equation, which, in this case, takes the form,
\begin{equation}\label{vlasov1}
p^r \partial_r F - \Gamma^j_{ab} p^a p^b \partial_{p^j}F = 0
\end{equation}
where $a, b$ correspond to $(t,r,\theta,\phi)$. It is understood in (\ref{vlasov1}) that $p^t$ is to be computed using $g_{ab}p^a p^b=-1$, so as to satisfy the ``mass shell restriction'' $\mu=1$, where $\mu$ is the particles mass. Therefore, in what follows we set,
\begin{equation}\label{p0}
p^t = \frac{1}{\sqrt{B(r)}}\sqrt{1+A(r) (p^r)^2 +h(r)^2 \left[ (p^{\theta})^2+\sin ^2 \theta (p^{\phi})^2\right]}
\end{equation}

We also notice that $p^t = d t/ d \tau$, where $\tau$ is proper time along the particle's world line. \\

The Einstein equations for the system are,
\begin{equation}\label{Einst1}
G_{ab} := R_{ab}-\frac{1}{2} R g_{ab} = 8 \pi T_{ab}
\end{equation}
with the energy momentum tensor given by,
\begin{equation}\label{Tmunu1}
T_{ab} = - \int{ F p_a p_b |g|^{1/2} \frac{dp^r dp^{\theta} dp^{\phi}}{p_t}}
\end{equation}
where $g$ is the determinant of $g_{ab}$, and $p_a= g_{ab} p^b$. Equations (\ref{vlasov1}), (\ref{Einst1}) and (\ref{Tmunu1}) define the Einstein-Vlasov system restricted to a static spherically symmetric space time, with the metric written in the form (\ref{one1}).

The assumption that the metric is static and spherically symmetric implies conservation of the particle's energy,
\begin{eqnarray}\label{energy1}
E & = & B(r) p^t \nonumber \\
 & = & \sqrt{B(r)}\sqrt{1+A(r) (p^r)^2 +h(r)^2 \left[ (p^{\theta})^2+\sin ^2 \theta (p^{\phi})^2\right]}
\end{eqnarray}
and of the square of its angular momentum per unit mass,
\begin{equation}\label{Lsquare}
L^2 = h(r)^4 \left[ \left(p^{\theta}\right)^2 + \sin^2 \theta \left(p^{ \phi} \right)^2 \right]
\end{equation}
It is easy to check that the Ansatz,
\begin{equation}\label{Ansatz01}
F(r,p^j) = \Phi(E,L^2)
\end{equation}
where $E$, and $L^2$ are the functions of $r$, and $p^j$ given by (\ref{energy1},\ref{Lsquare}), solves the Vlasov equation for an arbitrary function $\Phi$. \\

To construct and solve explicit models based on (\ref{Ansatz01}) for the metric (\ref{one1}), it is convenient to change integration variables in (\ref{Tmunu1}). We set,
\begin{eqnarray}
\label{change01}
  p^{\theta} &=& \frac{1}{h(r)^2} L \cos \chi \nonumber \\
  p^{\phi} &=& \frac{1}{h(r)^2 \sin \theta} L \sin \chi
\end{eqnarray}
and write (\ref{Tmunu1}) in the form,
\begin{equation}\label{Tmunu2a}
T_{ab}(r) =  \frac{1}{h(r)^2}
\sqrt{\frac{A(r)}{B(r)}}
\int{  \Phi(E,L^2) p_a p_b   \frac{L \;dL \; d\chi \; dp^r}{p^t}}
\end{equation}
where we should set,
\begin{eqnarray}
\label{Eandpt}
E &=&  \sqrt{B \left(1+ (p^r)^2 A + \frac{L^2}{h^2}\right)} \nonumber \\
  p^t &=& \sqrt{\frac{1}{B}\left(1+ (p^r)^2 A + \frac{L^2}{h^2}\right)}
\end{eqnarray}

Andreasson and Rein \cite{AndRein} have explored the properties of models where $\Phi$ takes the form,
\begin{equation}\label{andreassonrein}
\Phi(E,L^2)=\phi(E/E_0)(L^2-L_0^2)^{\ell}
\end{equation}

In this note we consider a different type of models, based on the Ansatz,
\begin{equation}\label{model01}
\Phi(E,L^2)=F(E)\;\Theta(E_0-E)\; \delta(L-L_0)
\end{equation}
where $\Theta(x)$ is the Heaviside step function, and $\delta(x)$ is Dirac's $\delta$; namely, we assume that $L$ takes only the single value $L_0$, and that there is an upper bound on $E$, given by $E_0$. $F(E) $ is assumed to be a smooth function of $E$. We then have,
\begin{eqnarray}
\label{Tmunu2}
  T_t{}^t &=& -\frac{4 \pi L_0 \sqrt{A}}{h^3} \int_0^{p^r_{max}}{F(E) \sqrt{h^2+L_0^2+(p^r)^2  h^2 A} dp^r}  \nonumber \\
  T_r{}^r &=& \frac{4 \pi L_0 \sqrt{A^3}}{h} \int_0^{p^r_{max}}{\frac{ F(E) (p^r)^2}{\sqrt{h^2+L_0^2+(p^r)^2  h^2 A}} dp^r} \nonumber \\
  T_{\theta}{}^{\theta} &=& \frac{ 2\pi (L_0)^3 \sqrt{A}}{h^3} \int_0^{p^r_{max}}{\frac{ F(E) }{\sqrt{h^2+L_0^2+(p^r)^2  h^2 A}} dp^r} \nonumber \\
T_{\phi}{}^{\phi} & = & T_{\theta}^{\theta}
\end{eqnarray}
where $p^r_{max}$ depends on $r$ and is given by,
\begin{equation}\label{a1}
p^r_{max}=\sqrt{\frac{1}{A(r)}}\sqrt{\frac{E_0^2}{B(r)}-1-\frac{L_0^2}{h(r)^2}}
\end{equation}
if $E_0^2 > B(r) (1+L_0^2/h^2)$, and $p^r_{\max}=0$ otherwise. This simply states the fact that $T_{ab} \neq 0$ only in those regions where a (test) particle with energy $E_0$ and angular momentum $L_0$ can actually move.

\section{Particular models}

We may now use the previous results to construct simple models and analyze their interpretation for a range of possible parameters. This analysis may be carried out in a number of ways. Here we choose the following; we first use a gauge freedom in the metric (\ref{one1}) to set,
\begin{eqnarray}
\label{two1}
  h(r) &=& r \nonumber \\
  A(r) &=& 1/(1-2m(r)/r)
\end{eqnarray}
Then, from the Einstein equations and the form (\ref{Tmunu2}) of $T_{ab}$, we find two independent equations for $m(r)$ and $B(r)$,
\begin{eqnarray}
\label{two2}
\frac{d m}{dr} &=& 4 \pi r^2 \rho(r) \nonumber \\
\frac{d B}{dr} &=& \frac{2 B(r)(m(r)+4 \pi r^3 p(r))}{r(r-2 m(r))}
\end{eqnarray}
where, $\rho(r)=-T_t{}^t$ is the energy density, and $p(r)=T_r{}^r$ is the radial pressure, given by (\ref{Tmunu2}), with $h(r)=r$. There is also an equation for $p_T(r)= T_{\theta}{}^{\theta}$, but, as can be checked, this is not independent of (\ref{two2}). Equations (\ref{two2}) are deceivingly simple, because the explicit dependence of $\rho$ and $p$ on $m$ and $B$ is in general quite complicated. Here we consider a simple example and propose a method for constructing the solutions, that is illustrated by the example. It can be seen that some simplification is attained if we choose,
\begin{equation}
\label{two3}
F(E) = Q_1 E =  Q1 \sqrt{B \left(1+ (p^r)^2 A + \frac{L^2}{r^2}\right)}
\end{equation}
where $Q_1 \geq 0$ is a constant. With this choice we may perform the integrals in (\ref{Tmunu2}) explicitly and, after some simplifications, we get,
\begin{eqnarray}
\label{two4a}
\frac{d m}{dr} &=&  \frac{Q_2 \left[2(L_0^2+r^2)B+r^2E_0^2\right]\sqrt{r^2 E_0^2 - (L_0^2+r^2)B}}{r^3 B} \\
\label{two4b}
\frac{d B}{dr} &=& \frac{2 m B}{r(r-2m )} +\frac{2Q_2\left[r^2 E_0^2- (L_0^2+r^2)B\right]^{3/2}}{r^3(r-2m)}
\end{eqnarray}
where $Q_2 = 16 \pi^2 L_0 Q_1/3$. We also find,
\begin{eqnarray}
\label{two5}
\rho(r) &=& \frac{Q_2 \left[2(L_0^2+r^2)B+r^2E_0^2\right]\sqrt{r^2 E_0^2 - (L_0^2+r^2)B}}{4 \pi r^5 B} \nonumber \\
p(r) &=& \frac{Q_2\left[r^2 E_0^2- (L_0^2+r^2)B\right]^{3/2}}{4 \pi B r^5} \nonumber \\
p_T(r) &=& \frac{3 Q_2 L_0^2 \sqrt{r^2 E_0^2- (L_0^2+r^2)B}}{8 \pi  r^5}
\end{eqnarray}
Considering (\ref{two4a},\ref{two4b}), we find that it is a simple, but rather difficult to handle, system of equations for $m(r)$ and $B(r)$. We have not found closed (analytical) solutions for the system, and, therefore, we must resort to numerical methods. The application of these methods requires, however, considering and solving several subtleties inherent in the system. As indicated above, in all these equations ((\ref{two4a},\ref{two4b}) and (\ref{two5})), the terms involving $Q_2$ should be set equal to zero if $r^2 E_0^2\leq (L_0^2+r^2)B$. We notice that for $Q_1=Q_2=0$ we have $m(r)= M$ with $M=\mbox{constant}$, and $B(r)=B_0(1-2 M/r)$, where $B_0$ is also a constant, corresponding to the standard Schwarzschild solution. For a shell type solution, these solutions correspond to the inner and outer regions, to be matched to the region where $T_{ab} \neq 0$. When $Q_1 \neq 0$, since we must have $B(r) > 0$, we must also have $dm/dr \geq 0$, but, even though $\rho \geq 0$ we might end up with $m(r) < 0$, and still have all equations satisfied.

For shell like solutions, either with an empty interior or with a central mass (black hole), a further difficulty can be seen considering that there should exist an ``allowed region'' where $r^2 E_0^2 \geq (L_0^2+r^2)B$, with $r$ taking values in the interval $r_i \leq r \leq r_o$, where $r_i$ and $r_o$ are, respectively, the inner and outer radii of the shell. We must impose continuity in both $B(r)$ and $m(r)$ to avoid $\delta$ functions in $T_{ab}$. This implies that $r^2 E_0^2- (L_0^2+r^2)B$ is continuous in $r_i \leq r \leq r_o$ and approaches continuously the value zero at the boundaries. Therefore, both $dm/dr$ and $dB/dr$ are also continuous inside and at the boundaries of this interval, and actually we have $dm/dr|_{r=r_i} =0$. We also find that $d^2B/dr^2$ should be continuous, but $d^2m/dr^2$ must be singular, and this makes the construction of numerical solutions where we try to fix from the beginning the values of $r_i$ and $r_o$ rather difficult. Nevertheless, the above analysis indicates that for $r > r_i$, but $r \sim r_i$, we should have,
\begin{eqnarray}
\label{two6}
B(r) &=& B_0 + B_1 (r-r_i) + B_2 (r-r_i)^2 + {\cal R}_B  \nonumber \\
m(r) &=& M_1+ {\cal R}_m
\end{eqnarray}
where $B_0$, $B_1$, and $M_1$ are constants and ${\cal R}_B$ and ${\cal R}_m$ are functions of  $r$ that vanish respectively faster than $(r-r_i)^2$ and $(r-r_i)$ for $r-r_i \to 0^+$. It is straightforward to extend this analysis to higher order by imposing consistency between the right and left hand sides of (\ref{two4a},\ref{two4b}) as $r-r_i \to 0^+$. We find,
\begin{eqnarray}
\label{two7}
B(r) &=& B_0 + B_1 (r-r_i) + B_2 (r-r_i)^2 + B_3 (r-r_i)^5/2+  \tilde{{\cal R}}_B  \nonumber \\
m(r) &=& M_1+ M_2 (r-r_i)^{3/2}+M_3 (r-r_i)^{5/2}+\tilde{{\cal R}}_m
\end{eqnarray}
where $B_2$, $B_3$, $M_2$ and $M_3$ are constants, and  $\tilde{{\cal R}}_B$ and $\tilde{{\cal R}}_m$ stand for higher order terms. The constants appearing in (\ref{two7}) are not independent. They may be written, e.g., in terms of $r_i$, $M_1$, $L_0$, $E_0$ and $Q_2$. We notice that $M_1$ is the Schwarzschild mass for the region inside the shell ($ r \leq r_i$). Moreover, the system (\ref{two4a},\ref{two4b}) is invariant under the the rescaling $B \to \lambda B$,  $E_0 \to \sqrt{\lambda} E_0$, and $Q_2 \to Q_2/\sqrt{\lambda}$.

The condition that $r=r_i$ corresponds to the inner boundary of the shell implies,
\begin{equation}\label{two8}
B_0 = \frac{r_i^2 E_0^2}{r_i^2+L_0^2}
\end{equation}
Similarly, we find,
\begin{eqnarray}
\label{two9}
B_1 &=& \frac{2 M_1 E_0^2 r_i}{(r_i-2M_1)(r_i^2+L_0^2)} \nonumber \\
B_2 &=& -\frac{ M_1 E_0^2 }{(r_i-2M_1)(r_i^2+L_0^2)} \nonumber \\
M_2 &=& Q_2 E_0 \sqrt{ \frac{8 (r_i^2+L_0^2)(r_iL_0^2-M_1r_i^2-3 M_1L_0^2)}{(r_i-2M_1)r_i^5}}
\end{eqnarray}
The explicit expressions for $B_3$ and $M_3$ are also easily obtained but are rather long and will not be included here, although they were used in the numerical computations described below.

\section{Some general properties}

An interesting question regarding the model of the previous Section is related to the possible values that the thickness of the shells can attain. This may be analyzed by considering the limit of solutions of the system (\ref{two4a},\ref{two4b}) as $r \to \infty$, under the restrictions that $\rho \neq 0$, and $r > 2 m(r)$.  The first, according to (\ref{two4a},\ref{two4b}) and  (\ref{two5}), implies $B(r) < E_0^2$, and $B_0=\lim_{r \to \infty}B(r) \leq E_0^2$. We remark that $dB/dr \geq 0$. Therefore, $B(r)$ must approach $B_0$ monotonically from below. Consider first the case $B_0 < E_0^2$. Replacing in the first equation in (\ref{two4a},\ref{two4b}), for large $r$ we find that $dm/dr$ approaches a constant value, and, therefore, $m(r)$ grows linearly with $r$. But then, replacing in the second equation in (\ref{two4a},\ref{two4b}), we find that $dB/dr$ decreases as $1/r$, leading to a logarithmic growth in $B(r)$, incompatible with the assumed conditions. Therefore, any possible solution should have $B_0=E_0^2$. Then, for large $r$ we should have,
\begin{equation}\label{three1}
B(r)=\frac{E_0^2 r^2}{L_0^2+r^2}-B_1(r)
\end{equation}
with $B_1(r)\to 0$, as $r\to \infty$. Replacing now in (\ref{two4a},\ref{two4b}), to leading order we find,
\begin{equation}\label{three2}
\frac{dm}{dr} \simeq 3 Q_2 \sqrt{B_1}
\end{equation}
and this implies that $m(r)/r \to 0$ as $r \to \infty$. Then, using again (\ref{two4a},\ref{two4b}), we should have,
\begin{equation}\label{three3}
\frac{dB_1}{dr} \simeq  - \frac{2E_0^3 Q_2}{r}
\end{equation}
and this implies $B_1(r) \simeq -2E_0^3Q_2 \ln(r)$, which contradicts the assumption $B_1(r) \to 0$. Thus we conclude that the equation,
\begin{equation}\label{three4}
E_0^2r^2-(L_0^2+r^2)B(r) = 0
\end{equation}
must always be satisfied for finite $r$, and, therefore, {\em all shells constructed in accordance with the prescription (\ref{two3}) have finite mass and finite thickness}.

We nevertheless believe that this results is more general, and applies to all shells satisfying the Ansatz (\ref{model01}), although we do not have a complete proof of this statement.

\section{Numerical results}

As indicated, we do not have closed solutions of the equations for $B(r)$ and $m(r)$, even for the simple model of the previous Section. Nevertheless, since (\ref{two4a},\ref{two4b}) is a first order ODE system, we can apply numerical methods to analyze it. We may use the expansions (\ref{two7}), (disregarding the terms in $\tilde{{\cal R}}_B$ and $\tilde{{\cal R}}_m$), to obtain appropriate initial values for $B(r)$ and $m(r)$, for $r$ close to $r_i$, in the non trivial region $r > r_i$.

We may illustrate this point with a particular example. We take $r_i = 7.0$, $M_1=1.0$, $L_0=4$, $E_0=1$ and $Q_2=0.1$. Using these values and (\ref{two7}) (truncated as indicated above), we find $B(7.0001)=0.7538...$, and $m(7.0001)=1.0000...$ (actually, the computations were carried out to 30 digits, using a Runge-Kutta integrator). The numerical results are plotted in Figure 1 and Figure 2.

\begin{figure}
\centerline{\includegraphics[height=12cm,angle=-90]{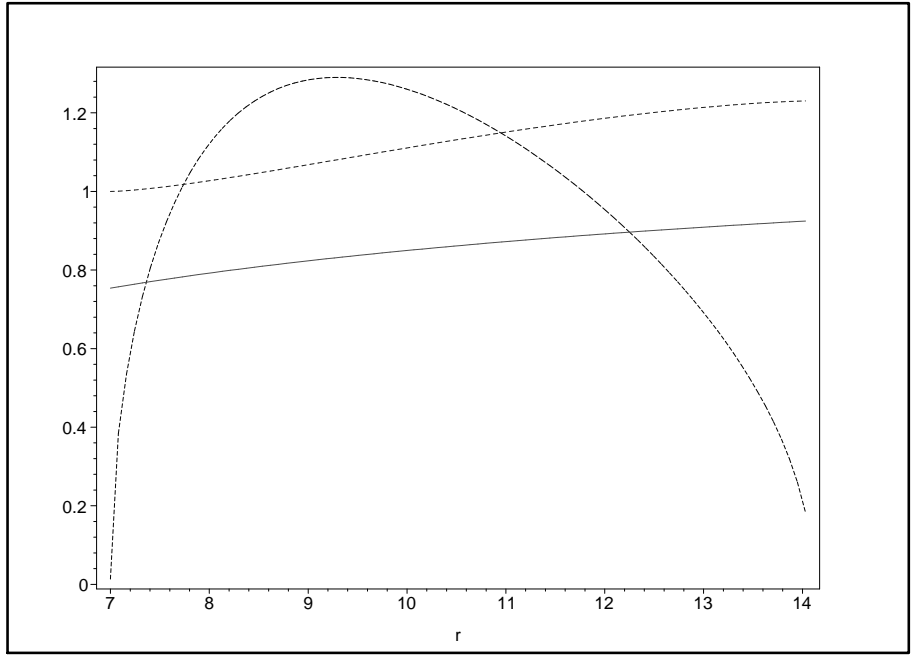}}
\caption{Plots of $B(r)$ (solid), $m(r)$ (dotted), and $30 (dm/dr)$ (dashed), as functions of $r$ for $r_i = 7.0$, $M_1=1.0$, $L_0=4$, $E_0=1$ and $Q_2=0.1$. The end point of the plot is at $r_o=14.1098...$, with $B(r_o)=0.9256...$ and $M_2=m(r_o)=1.2306...$}
\end{figure}

\begin{figure}
\centerline{\includegraphics[height=12cm,angle=-90]{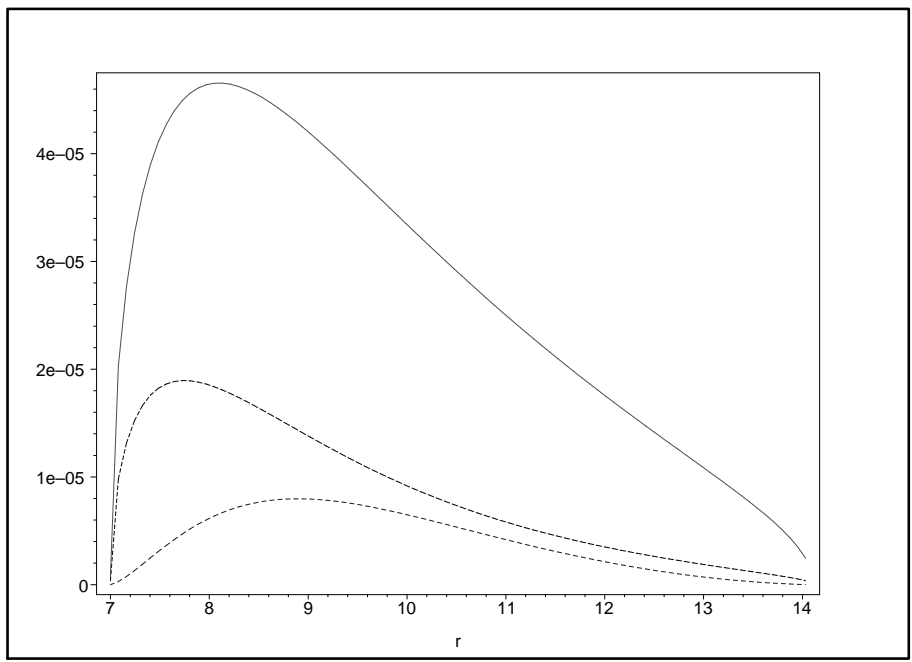}}
\caption{Plots of $\rho(r)$ (solid, higher curve), $4 p_T(r)$ (dashed, middle curve), and $40  p(r)$ (dotted, lower curve), as functions of $r$ for $r_i = 7.0$, $M_1=1.0$, $L_0=4$, $E_0=1$ and $Q_2=0.1$. The end point of the plot is at $r_o=14.1098...$, with $B(r_o)=0.9256...$ and $M_2=m(r_o)=1.2306...$}
\end{figure}

\section{Thin shells}

One of the motivations for studying the type of shells considered in this paper is the possible existence of a non trivial ``thin shell'' limit, where the thickness of the shell goes to zero, with the restriction to a single or a finite set of values of $L$, and how this limit compares with the thin shells considered in \cite{Evans} and \cite{gleram}. We remark that the existence of thin shell limits of Einstein-Vlasov systems has already been analyzed in the literature \cite{andreassonCMP}. Here we are interested not only in the existence of this limit for our particular models, but especially in the limiting values of the parameters characterizing our shells. Since this type of analysis is not immediately included in, e.g, \cite{andreassonCMP}, we consider it relevant to provide an explicit proof of the properties of our models in the thin shell limit.

We first recall that for a static thin shell constructed according to Evans' prescriptions \cite{Evans}, we have the following relation between the radius $R$, inner ($M_1$) and outer  ($M_2$) mass, and angular momentum $\tilde{L}_0$ of the particles,
\begin{equation}\label{five1}
\tilde{L}_0^2=\frac{R-\sqrt{R-2M_1}\sqrt{R-2M_2}}{3\sqrt{R-2 M_1}\sqrt{R-2M_2}-R}
\end{equation}

We may now prove that the non trivial thin shell limits of the shells constructed according to the prescription (\ref{two3}) effectively coincide with the Evans shells of reference \cite{Evans} as follows. We first take the $r$ derivative of (\ref{two4b}), and then use (\ref{two4a}) and (\ref{two4b}, to obtain,
\begin{eqnarray}
\label{thin01}
\frac{d^2B}{dr^2} &=& \frac{\left[ r(2 E^2 r^2 +(r^2+L^2)B)\dfrac{dB}{dr}+2(4E^2r^2+(2L^2-r^2)B)\right]}{r(E^2 r^2 +2(L^2+r^2)B)(r-2m)}\frac{dm}{dr} \nonumber \\
 & &  -\frac{4 }{r^2(r-2m)}\left[r(r-2m)\frac{dB}{dr}-m B\right]
\end{eqnarray}

Next let $r_i$ and $r_o$ be, respectively, the inner and outer radii of the shell, and $M_1=m(r_i)$ and $M_2=m(r_o)$ the corresponding masses inside and outside the shell. Then, for $r_i \leq r \leq r_o$ we have $dB/dr >0$, and,
\begin{equation}\label{thin02}
\left.\frac{dB}{dr}\right|_{r=r_i} = \frac{2M_1 B(r_i)}{r_i(r_i-2M_1)} \;\;;\;\;\left.\frac{dB}{dr}\right|_{r=r_o} = \frac{2M_2 B(r_o)}{r_o(r_o-2M_2)}
\end{equation}

The idea now is to use the fact that,
\begin{eqnarray}
\label{thin03}
  \int_{r_i}^{r_o}{\frac{d^2B}{dr^2}} dr &=&  \left.\frac{dB}{dr}\right|_{r=r_o}-\left.\frac{dB}{dr}\right|_{r=r_i} = \frac{2M_2 B(r_o)}{r_o(r_o-2M_2)} -\frac{2M_1 B(r_i)}{r_i(r_i-2M_1)} \; ;\nonumber \\
 \int_{r_i}^{r_o}\frac{dm}{dr} dr &=& M_2-M_1
\end{eqnarray}

On this account we rewrite (\ref{thin01}) in the form,
\begin{eqnarray}
\label{thin04}
\frac{1}{r-2m}\frac{dm}{dr} &=& \frac{r (E^2 r^2+2B(L^2+r^2))}{\left[ r(2 E^2 +(r^2+L^2)B)\dfrac{dB}{dr}+2(4E^2r^2+(2L^2-r^2)B)\right]}\frac{d^2B}{dr^2}
 \nonumber \\
 & &  +\frac{4 (E^2 r^2+2B(L^2+r^2))\left[r(r-2m)\frac{dB}{dr}-m B\right]}
 {r (r-2m)\left[ r(2 E^2 +(r^2+L^2)B)\dfrac{dB}{dr}+2(4E^2r^2+(2L^2-r^2)B)\right]}
\end{eqnarray}
and integrate both sides from $r=r_i$ to $r=r_o$. But now we notice that while both $m(r)$ and $dB/dr$ are rapidly changing but bounded in $r_i \leq r \leq r_o$, the change of $B(r)$, and $r$ in that interval is only of order $r_o-r_i$. We may then choose a point $r=R$, with $r_i < R < r_o$, and set $B(r)=B(R)=B_0$, and $r=R$, except in the arguments of $m(r)$, $dm/dr$, $dB(r)/dr$ and $d^2 B(r)/dr^2$, in (\ref{thin04}), as this introduces errors at most of order $r_o-r_i$, in the factors of $dm/dr$, and $d^2 B(r)/dr^2$, and in the last term in the right hand side of (\ref{thin04}). Similarly, we may set $E^2=(L^2+R^2)B_o/R^2$ in (\ref{thin04}), to obtain, up to terms of order $r_o-r_i$,
\begin{eqnarray}
\label{thin05}
\frac{1}{R-2m(r)}\frac{dm}{dr} &=& \frac{R (L^2+R^2))}{  R(R^2+L^2))\dfrac{dB}{dr}+2(R^2+2L^2)B_0}\frac{d^2B}{dr^2}
 \nonumber \\
 & &  +\frac{4 ( (L^2+R^2))\left[R(R-2m)\frac{dB}{dr}-m B_0\right]}
 {R (R-2m)\left[R(R^2+L^2)\dfrac{dB}{dr}+2(R^2+2 L^2)B_0\right]}
\end{eqnarray}
and, again, we notice that the last term on the right of (\ref{thin05}) gives a contribution of order $r_o-r_i$. We then conclude that,
\begin{equation}\label{thin06}
\lim_{r_o\to r_i} \int_{r_i}^{r_o}\frac{1}{R-2m(r)}\frac{dm}{dr} dr =
\lim_{r_o\to r_i} \int_{r_i}^{r_o}\frac{R (L^2+R^2))}{  R(R^2+L^2))\dfrac{dB}{dr}+2(R^2+2L^2)B_0}\frac{d^2B}{dr^2} dr
\end{equation}
The integration of the terms in $dm/dr$, and $d^2 B(r)/dr^2$, is now straightforward. We use next (\ref{thin02}) and the fact that in this limit $B(r_i)=B_0=B(r_o)$ and $r_i=R=r_o$, to obtain,
\begin{equation}\label{thin07}
\frac{\sqrt{R-2M_1}}{\sqrt{R-2M_2}} = \frac{(R-2M_1)(R^3+2RL^2-3M_2L^2-M_2R^2)}{(R-2M_2)(R^3+2RL^2-3M_1L^2-M_1R^2)}
\end{equation}

Solving this equation for $L^2$, we finally find,
\begin{equation}\label{thin08}
L^2 = \frac{R^2 (R-\sqrt{R-2M_1}\sqrt{R-2M_2})}{3\sqrt{R-2M_1}\sqrt{R-2M_2}-R}
\end{equation}
which is, precisely, the relation satisfied by the parameters of the shells of (\ref{five1}).

We can also check this result, and, in turn, the accuracy of numerical codes, by directly considering initial data for the numerical integration that effectively lead to shells where the thickness is a small fraction of the radius.

\begin{figure}
\centerline{\includegraphics[height=12cm,angle=-90]{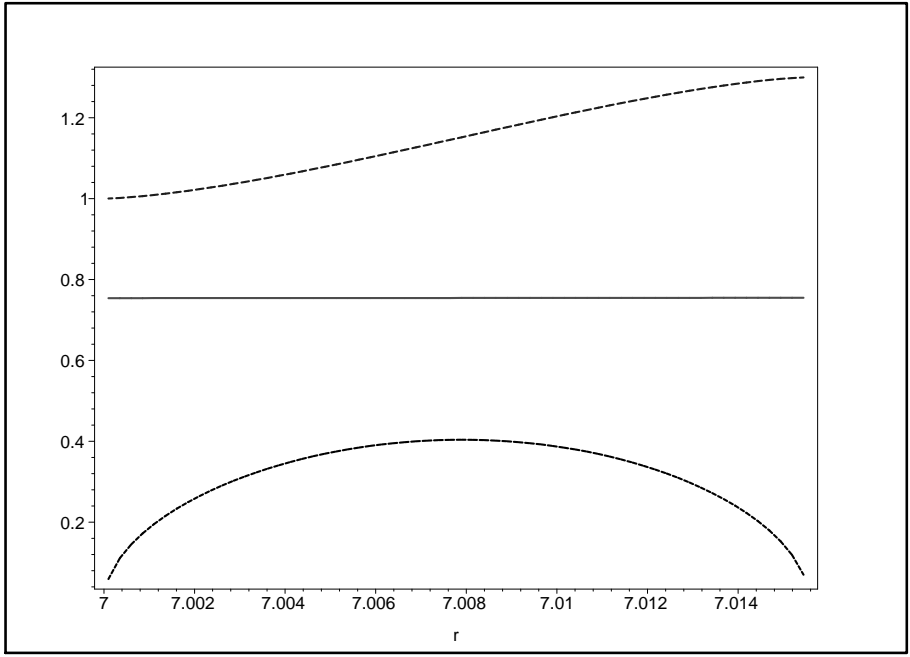}}
\caption{Plots of $m(r)$ (dotted, higher curve), $B(r)$ (solid, middle curve),   and $10 \rho$ (dashed, lower curve), as functions of $r$ for $r_i = 7.0$, $M_1=1.0$, $L_0=4$, $E_0=1$ and $Q_2=800$. The end point of the plot is at $r_o=7.0155...$, with $B(r_o)=0.7546...$ and $M_2=m(r_o)=1.2999...$}
\end{figure}

A particular example is given in Figure 3, where the values of the initial data is also indicated. We can see that the mass increases by about $30$ percent, while the thickness of the shell is less than $0.3$ percent of the shell radius. We can check that these results are in agreement with (\ref{five1}).
Solving this for $M_2$,
\begin{equation}\label{five2}
M_2=\frac{R\left[2(R-3M_1)R^2 \tilde{L}_0^2+(4R-9M_1) \tilde{L}_0^4-M_1R^4\right] }{(R-2M_1)(R^2+3 \tilde{L}_0^2)^2 }
\end{equation}
and replacing $\tilde{L}_0=4$, $M_1=1$, and $R=7$, we find $M_2=1.2997...$ in very good agreement with the numerical results quoted in Figure 3.

\section{Stability analysis}

Consider a shell approaching the thin shell limit. We restrict to the case of vanishing inner mass $M_1=0$. If $r_i$ and $r_o$ are, respectively the inner and outer boundaries of the shell, the matching conditions in the absence of singular shells at $r_i$ and $r_o$ imply that both $m(r)$ and $B(r)$ and their first derivatives should be continuous at both $r=r_i$ and $r=r_o$. The form (\ref{one1}) of the metric with the choice (\ref{two1}) imply that for $0 \leq r \leq r_i$ we have,
\begin{equation}\label{six1}
ds^2= -B_i dt^2 + dr^2 + r^2 d \Omega^2
\end{equation}
where $B_i >0$ is a constant. Then, at $r=r_i$ we have,
\begin{equation}\label{six2}
B_i= \frac{r_i^2 E_0^2}{L_0^2 +r_i^2}
\end{equation}

For $r\geq r_o$, the metric takes the form,
\begin{equation}\label{six3}
ds^2= -B_o  \left( \frac{1-\frac{2M_2}{r}}{1-\frac{2M_2}{r_o}}\right)dt^2 +\left( 1-\frac{2M_2}{r}\right)^{-1} dr^2 + r^2 d \Omega^2
\end{equation}
where $B_o>0$ is another constant satisfying,
\begin{equation}\label{six4}
B_o= \frac{r_o^2 E_0^2}{L_0^2 +r_o^2}
\end{equation}

Consider now a test particle moving along a geodesic of the shell space time, with 4-velocity $ U^{\mu}(\tau) = (dt/d\tau,dr/d\tau,dt\theta/d\tau,d\phi/d\tau)$, with $U^{\mu} U_{\mu}=-1$. Without loss of generality we may choose $d\theta/d\tau=0$. Then we have the constants of the motion:
\begin{equation}\label{six5}
E= B(r) \frac{dt}{d\tau} \;\;\;,\;\;\; L=r^2\frac{d\phi}{d\tau}
\end{equation}
and the normalization of $U^{\mu}$ implies,
\begin{equation}\label{six6}
\left(\frac{dr}{d\tau}\right)^2= \frac{E^2}{B(r)}\left( 1-\frac{2m(r)}{r}\right)-\left(1+\frac{L^2}{r^2}\right)\left( 1-\frac{2m(r)}{r}\right)
\end{equation}
Therefore, for $r \geq r_o$ we have,
\begin{equation}\label{six7}
\left(\frac{dr}{d\tau}\right)^2 = \frac{E^2}{B_o}\left(1-\frac{2M_2}{r_o}\right) -\left(1+\frac{L^2}{r^2}\right)\left( 1-\frac{2M_2}{r}\right)
\end{equation}
and the particle radial acceleration is given by,
\begin{equation}\label{six8}
\frac{d^2r}{d\tau^2}= \frac{L^2}{r^3}-\frac{(3L^2+r^2)M_2}{r^4}
\end{equation}
If we assume now that the shell is close to the thin shell limit, with angular momentum $L_0$, radius $R$, (with $r_i < R < r_o$), and mass $M_2$, then we should have,
\begin{equation}\label{six9}
L_0^2= \frac{(\sqrt{R} -\sqrt{R-2M_2}) R^2}{3\sqrt{R-2M_2}-\sqrt{R}}
\end{equation}
Then, for $L\simeq L_0$, and $r\simeq R\simeq r_o$ we find,
\begin{equation}\label{six10}
\frac{d^2r}{d\tau^2}\sim - \frac{2 \sqrt{R-2M_2}(\sqrt{R}-\sqrt{R-2M_2})}{R^{3/2}(3\sqrt{R-2M_2}-\sqrt{R})} < 0
\end{equation}

Similarly, in the region $0 \leq r\leq r_i$, for $r \sim r_i \sim R$ we find,
\begin{equation}\label{six11}
\frac{d^2r}{d\tau^2}\sim  \frac{2  (\sqrt{R}-\sqrt{R-2M_2})}{R (3\sqrt{R-2M_2}-\sqrt{R})} > 0
\end{equation}
and, therefore, all these shells are stable under ``single particle evaporation'', in total agreement with the results of \cite{gleram}.

A related problem is that of the dynamical stability of the shell as a whole, as was also analyzed in \cite{gleram}. There, the shells considered where ``thin'', and therefore, the motion was described by ordinary differential equations for the shell radius $R$ as a function, e.g., of proper time on the shell, which allowed for a significant simplification of the treatment of the small departures from the equilibrium configurations. Unfortunately, the corresponding equations of motion for the shells considered here would be considerably more complicated and their analysis completely outside the scope of the present research. We, nevertheless, expect that such treatment, if appropriately carried out, would also agree with the results found in \cite{gleram} as the ``thin shell'' limit is approached.

\section{Shells with two or more values of the angular momentum}

It is rather simple to extend the analysis of the previous Sections to the case where the angular momentum of the particles takes on a discrete, finite set of values. Instead of (\ref{model01}), we have,
\begin{equation}\label{model201}
\Phi(E,L^2)=\sum_i F_i(E)\;\Theta(E_i-E)\; \delta(L-L_i)
\end{equation}
where the functions $F_i \geq 0$ are arbitrary, with $i=1,2,...,N$, and $N$  finite. We will restrict to the case of two separate values, (N=2), since, as will be clear from the treatment, the extension to a larger number of components is straightforward. We are actually interested in the behaviour of these shells as they approach a common thin shell limit. Therefore, we will further simplify our Ansatz to the form,
\begin{equation}
\label{twodos3}
F_i(E) = Q_i E =  Q_i \sqrt{B \left(1+ (p^r)^2 A + \frac{L_i^2}{r^2}\right)}
\end{equation}
where $Q_i \geq 0$ are constants. With this choice we may perform the integrals in (\ref{Tmunu2}) explicitly and, after some simplifications, we get,
\begin{eqnarray}
\label{twodos4}
\frac{d m}{dr} &=&
\frac{C_1 \left[2(L_1^2+r^2)B+r^2E_1^2\right]\sqrt{r^2 E_1^2
- (L_1^2+r^2)B}}{r^3 B} \nonumber \\
& & +
\frac{C_2 \left[2(L_2^2+r^2)B+r^2E_2^2\right]\sqrt{r^2 E_2^2
- (L_2^2+r^2)B}}{r^3 B}
\nonumber \\
\frac{d B}{dr} &=& \frac{2 m B}{r(r-2m )} \nonumber \\
& & +
\frac{2C_1\left[r^2 E_1^2- (L_1^2+r^2)B\right]^{3/2}}{r^3(r-2m)}
+\frac{2C_2\left[r^2 E_2^2- (L_2^2+r^2)B\right]^{3/2}}{r^3(r-2m)}
\end{eqnarray}
where $C_i = 16 \pi^2 L_i Q_i/3$, $i=1,2$. We also find,
\begin{eqnarray}
\label{twodos5}
\rho(r) &=& \frac{C_1 \left[2(L_1^2+r^2)B+r^2E_1^2\right]\sqrt{r^2 E_1^2 - (L_1^2+r^2)B}}{4 \pi r^5 B} \nonumber \\
& & +\frac{C_2 \left[2(L_2^2+r^2)B+r^2E_2^2\right]\sqrt{r^2 E_2^2 - (L_2^2+r^2)B}}{4 \pi r^5 B} \nonumber \\
p(r) &=& \frac{C_1\left[r^2 E_1^2- (L_1^2+r^2)B\right]^{3/2}}{4 \pi B r^5}
+\frac{C_2\left[r^2 E_2^2- (L_2^2+r^2)B\right]^{3/2}}{4 \pi B r^5} \nonumber \\
p_T(r) &=& \frac{3 C_1 L_1^2 \sqrt{r^2 E_1^2- (L_1^2+r^2)B}}{8 \pi  r^5}
+\frac{3 C_2 L_2^2 \sqrt{r^2 E_2^2- (L_2^2+r^2)B}}{8 \pi  r^5}
\end{eqnarray}

It is clear that we recover the results of the previous Sections if we set either $C_1$ or $C_2$ equal to zero.
\\

It will be convenient to define separate contributions to the density, $\rho_1$ and $\rho_2$, for the particles with $L_1$ and $L_2$.
\begin{eqnarray}
\label{rho12}
\rho_1(r) &=& \frac{C_1 \left[2(L_1^2+r^2)B+r^2E_1^2\right]\sqrt{r^2 E_1^2 - (L_1^2+r^2)B}}{4 \pi r^5 B} \nonumber \\
\rho_2(r) &=& \frac{C_2 \left[2(L_2^2+r^2)B+r^2E_2^2\right]\sqrt{r^2 E_2^2 - (L_2^2+r^2)B}}{4 \pi r^5 B}
\end{eqnarray}
Then, provided the integrations cover the supports of both $\rho_1$ and $\rho_2$, we have,
\begin{equation}\label{deltam1}
M_2-M_1 = \Delta M = \Delta m_1 +\Delta m_2
\end{equation}
where
\begin{equation}\label{deltam2}
 \Delta m_1 = \int{ 4 \pi r^2 \rho_1 dr}\;\; , \;\;\ \Delta m_2= \int{ 4 \pi r^2 \rho_2 dr}
\end{equation}
These expressions may be considered as the contributions to the mass from each class of particles. This will be used in the next section to compare numerical results with the thin shell limit.

Equations (\ref{twodos4}) may be numerically solved for appropriate values of the constants $C_i$, $E_i$, and $L_i$, and initial values, i.e., for some $r$, of $m(r)$ and $B(r)$. We remark that, as in the previous Sections, it is understood in (\ref{twodos4}), (and also in (\ref{twodos5})), that both $\sqrt{ r^2 E_i^2- (L_i^2+r^2)B }$ and  $\left[r^2 E_i^2- (L_i^2+r^2)B\right]^{3/2}$ must be set equal to zero for $  r^2 E_i^2\leq  (L_i^2+r^2)B$. In this general case, it is clear that the shells (where by a ``shell'' we mean here the set of particles having the same angular momentum $L_i$) may be  completely separated or they may overlap only partially. We are particularly interested in the limit of a common thin shell for the chosen values of $L_i$. One way of ensuring that at least one of the shells completely overlaps the other is the following. We choose an inner mass $M_1$, and an inner radius $r_i$. This implies $m(r_i)=M_1$, while $B(r_i)=B_i$ is arbitrary. If we choose now arbitrary values for $L_1$ and $L_2$, the density $\rho$ will vanish at $r=r_i$ if we choose,
\begin{eqnarray}
\label{twodos6}
E_1^2 &=& \frac{B_i(L_1^2+r_i^2)}{r_i^2}  \nonumber \\
E_2^2 &=& \frac{B_i(L_2^2+r_i^2)}{r_i^2}
\end{eqnarray}
Actually we also need to impose,
\begin{eqnarray}
\label{twodos7}
L_1^2 &>& \frac{ r_i^2 M_1}{r_i-3M_1}  \nonumber \\
L_2^2 &>& \frac{ r_i^2 M_1}{r_i-3M_1}
\end{eqnarray}
to make sure that $r=r_i$ is the {\em inner} and not the {\em outer} boundary of the shells. We shall assume from now on that $L_2 > L_1$. Since, using the same arguments as in the single shell case, the shells have finite extension, it follows that one of the shells will be completely contained in the other.

In the next Section we display some numerical results, both for thick shells that overlap partially, and for shells approaching the thin shell limit. We again find that the limit is associated to large values of the $C_i$, and that the parameters describing the shells approach the thin shell values found in \cite{gleram}.

\section{Numerical results for two component systems}

As a first example we take $L_1=5$, $L_2=6.5$, $r_i=7$, and $M_1=1$. We also set $B(r_i)=1$, for simplicity. Then, from (\ref{twodos6}), we set $E_1^2= 1.510...$, $E_2^2=1.862..$. Finally, we choose $C_1=5$, $C_2=3$, and carry out the numerical integration. The results obtained indicate that the particles with $L_1=5$ are contained in the region $7.0 \leq r \leq 7.2682..$, while for $L_2 =6.5$ the corresponding range is $ 7.0 \leq r \leq 7.4764...$. The resulting value of the external mass in $M_2=2.1338...$. In Figure 4 we display the total density $\rho$ as a function of $r$ (solid curve), as well as the contributions $\rho_1$ and $\rho_2$ to the density from the particles with respectively $L_1=5$ (dashed curve) and with $L_2=6.5$ (dotted curve).

\begin{figure}
\centerline{\includegraphics[height=12cm,angle=-90]{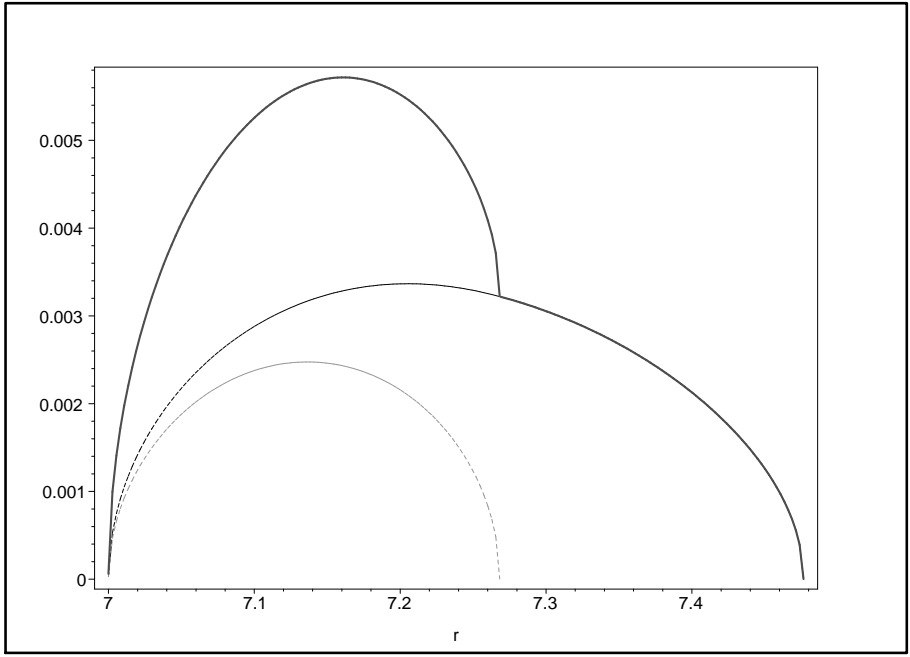}}
\caption{Plots of $\rho(r)$ (solid), $\rho_1(r)$ (dotted), and $\rho_2(r)$ (dashed), as functions of $r$ for $r_i = 7.0$, $M_1=1.0$, $L_1=5$, $L_2=6.5$, $E_1^2= 1.51...$, $E_2^2=1.86..$  and $C_1=5$, $C_2=3$. The end point of the plot is at $r_o=7.4764...$, with $B(r_o)=1.0606...$ and $M_2=m(r_o)=2.1338...$}
\end{figure}

As an illustration of the approach to the thin shell configuration we considered again the previous values $L_1=5$, $L_2=6.5$, $r_i=7$, $M_1=1$, $B(r_i)=1$, but choose $C_1=800$, $C_2=240$, and carried out the numerical integration. Figure 5 displays the functions $\rho(r)$ (solid curve), $\rho_1(r)$ (dashed curve), and $\rho_2(r)$ (dotted curve). We see that now the shell extends only to the region $r_i = 7.0 \leq r \leq r_o=7.020...$, i.e., its thickness is less than one percent of its radius. The mass, on the other hand, increases by roughly a factor of two, since $M_2=m(r_o)=2.022...$.

\begin{figure}
\centerline{\includegraphics[height=12cm,angle=-90]{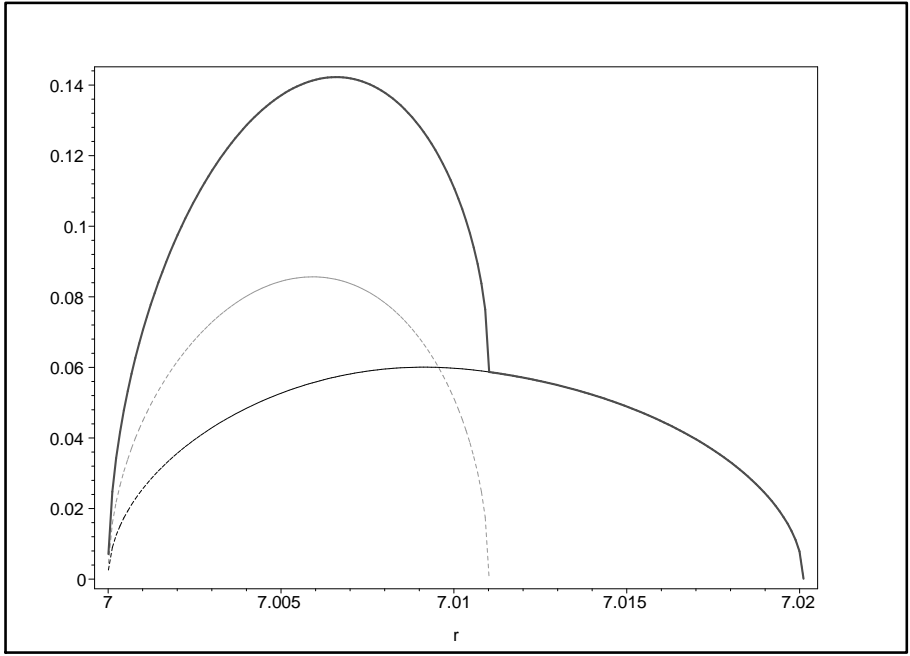}}
\caption{Plots of $\rho(r)$ (solid), $\rho_1(r)$ (dotted), and $\rho_2(r)$ (dashed), as functions of $r$ for $r_i = 7.0$, $M_1=1.0$, $L_1=5$, $L_2=6.5$, $E_1^2=1.51..$, $E_2^2=1.86..$ and $C_1=800$, $C_2=240$. The end point of the plot is at $r_o=7.020...$, with $B(r_o)=1.0026...$ and $M_2=m(r_o)=2.022...$}
\end{figure}

We may compare these results with those of the thin shell limit of \cite{gleram} as follows. It can be seen from (29) and (31) in \cite{gleram} that for a thin shell of radius $R$, inner mass $M_1$ and outer mass $M_2$, with two components with angular momenta $L_1$, and $L_2$, the ratio of the contributions of each component to the total mass is given by,
\begin{equation}\label{ratio1}
\frac{\Delta m_2}{\Delta m_1} = \frac{(\tilde{L}_0^2-L_1^2)(R^2+L_2^2)}{(L_2^2-\tilde{L}_0^2)(R^2+L_1^2)}
\end{equation}
where $\tilde{L}_0$ is given by (\ref{five1}). The numerical integration gives $\Delta m_1= 0.4545...$, $\Delta m_2 =0.5708...$. If now take $R=7$, and solve (\ref{ratio1}) for $\tilde{L}_0$ we find, $\tilde{L}_0=5.8079...$, while replacement in (\ref{five1}) gives $\tilde{L}_0=5.8386...$, which we consider as a good agreement, with a discrepancy of the order of the ratio of thickness to radius.

\section{Final comments and conclusions}

The general conclusion from this work is that one can effectively construct a wide variety of models satisfying the restriction that $L$ takes only a finite set of values, and that they do seem to contain the models used in \cite{gleram} as appropriate thin shell limits. We remark also that the starting point for our construction is a variant of the Ansatz used in \cite{AndRein}, where $f$ is factored in an $E$ (the particle energy) and an $L$ dependent terms. The possibility of multi-peaked structure in the case of more than one value of $L$ obtained here is also in correspondence with the general results obtained in \cite{AndRein}.


\section*{Acknowledgments}

This work was supported in part by grants from CONICET (Argentina)
and Universidad Nacional de C\'ordoba.  RJG and MAR are supported by
CONICET. We are also grateful to H. Andreasson for his helpful comments.

\end{document}